\documentclass[aip,twocolumn, floatfix,showpacs]{revtex4}
\usepackage{epsfig}
\usepackage{amsmath, amsthm, amssymb}
\usepackage[colorlinks=true,linkcolor=red, citecolor=blue]{hyperref}
\usepackage{graphicx}
\usepackage{hypernat}
\newcommand{\citebyname}[1]{~\citeauthor{#1}~\cite{#1}}
\begin{document}

\title{First principle electronic, structural, elastic, and optical properties of strontium titanate}

 \author{Chinedu E. Ekuma$^1$}
\author{Mark Jarrell$^1$}
\author{Juana Moreno$^1$}
\author{Diola Bagayoko$^2$}
\affiliation{$^1$Department of Physics and Astronomy \&
Center for Computation and Technology,
Louisiana State University,
Baton Rouge, Louisiana 70803, USA \\
$^2$Department of Physics, Southern University and 
A\&M College, Baton Rouge, Louisiana 70813, USA}

\date{\today}

\begin{abstract}
We report self-consistent ab-initio electronic, structural, elastic, 
and optical properties of cubic SrTiO$_{3}$ perovskite. Our 
non-relativistic calculations employed a generalized gradient approximation 
(GGA) potential and the linear combination of atomic 
orbitals (LCAO) formalism. The distinctive feature of our computations stem from 
solving self-consistently the system of equations 
describing the GGA, using the Bagayoko-Zhao-Williams (BZW) method. Our results are 
in agreement with experimental ones where 
the later are available. In particular, our theoretical, indirect band gap of 3.24 eV, 
at the experimental lattice constant of 
3.91 \AA{}, is in excellent agreement with experiment. Our predicted, equilibrium 
lattice constant is 3.92 \AA{}, with 
a corresponding indirect band gap of 3.21 eV and bulk modulus of 183 GPa.  
\end{abstract}

\pacs{71.15.Mb, 71.20.-b, 71.20.-b, 71.20.Mq, 71.20.Nr}
\maketitle

\section{Introduction}
\label{sec:intro}
Strontium titanate (SrTiO$_{3}$) is one of the most studied oxides of the ABO$_{3}$ 
perovskite type structures, due to its great 
technological importance. Many interesting phenomena such as 
colossal magnetoresistance, high-$T_{c}$ superconductivity, 
multiferroicity, and ferroelectricity are observed in complex 
oxides. Since most of the interesting complex oxides have perovskite 
structure, SrTiO$_{3}$ is an ideal starting point for their study. 
It has been widely used for integration with 
other oxides into heterostructures. Those heterostructres 
show interesting properties such as thermoelectricity\cite{Ohta2007,Wunderlich2009} and 
superconductivity\cite{Caviglia2008,Chang2010}.
Many new concepts of modern condensed matter and the physics 
of phase transitions have been developed while investigating 
this unique material\cite{Heifets2006, Wang2001, Lines1997}. 
SrTiO$_{3}$ has applications in the fields of ferroelectricity, optoelectronics 
and macroelectronics. It is used as a substrate for the epitaxial 
growth of high temperature superconductors. SrTiO$_{3}$ exhibits 
a very large dielectric constant. In comparison with SiO$_{2}$, 
SrTiO$_{3}$ has almost two orders of magnitude higher dielectric 
constant and may as well offer a better replacement for SiO$_{2}$ 
in Si-based nanoelectronic devices (see \citet{Wilk2001}[and references therein]). SrTiO$_{3}$ 
has found usage in optical switches, grain-boundary barrier layer capacitors, catalytic 
activators, waveguides, laser frequency doubling, high capacity 
computer memory cells, oxygen gas sensors, semiconductivity, etc
\cite{Eglitis2004, piskunov2004, Tinte1998, Jiangni2006, Cai2004, Bednorz1984, Balachandran1981, Kim1985}. 

During the last few decades, the electronic, structural, elastic, and optical properties of 
SrTiO$_{3}$ (STO), as a model of ABO$_{3}$ perovskite, have been under intensive investigation both 
experimentally\cite{Bickel1989, Maus2002, Charlton2000, Ikeda1999, Reihl1984, Pertosa1978, Brookes1986} and theoretically 
\cite{Eglitis2004, piskunov2004, Tinte1998, Jiangni2006, Cai2004, Padilla1997, Padilla1998, Cheng2000, Tinte2000}. 
But, from a theoretical point of view, a proper description of its electronic 
properties is still an area of active research. Theoretical computations have had difficulty in predicting 
the correct band gap energy and other related electronic properties of SrTiO$_{3}$ from first principle. 
The density functional theory plus additional Couloumb interactions (DFT+U) formalism 
\cite{Anisimov_a1991, Anisimov1991, Anisimov1997, Madsen2005} has had good 
successes in obtaining correct energy bands and gaps of materials, but can only be applied to 
correlated and localized electrons, e.g., 3$\textit{d}$ or 4$\textit{f}$ in transition and rare-earth oxides. 
The hybrid functionals (for e.g., Heyd-Scuseria-Ernzerhof (HSE)  hybrid functional \cite{Heyd2006,Heyd2004,Heyd2003}) 
has also been used in attempt to 
improve on the energy bands and band gaps of materials. This approach involves 
a range separation of the exchange energy into some fraction of 
nonlocal Hartree-Fock exchange potential and a fraction of local spin density approximation (LSDA) or 
generalized gradient approximation (GGA) exchange potential. We should note that this range separation is not 
universal. There is always a range separation parameter $\omega$ which varies between 0 and $\infty$.  
While it is reasonably clear that there exists a value
of $\omega$ that gives the correct gap for a given system, 
this $\omega$ is not universal as it is always adjusted from one system to another \cite{PhysRevB.78.121201,Henderson20011}. 
For example, in HSE06 \cite{Heyd2006,Krukau2006}, 
$\omega$ $=$ 0.11$a_{0}^{-1}$ ($a_{0}$ 
is the Bohr radius) and in Perdew-Burke-Ernzerhof (PBEh) global hybrid \cite{Janesko2008}, it is 25 $\%$ 
short-range exact exchange and 75 $\%$ short-range 
PBE exchange. Even though the HSE functional, in most cases, accurately 
reproduces the optical gap in semiconductors, it severely underestimates 
the gap in insulators \cite{Henderson20011,PhysRevB.76.195440} and its band width
in metallic systems is generally too large \cite{Henderson20011,PhysRevB.76.195440,Paier2006,PhysRevB.83.195134}. 
The \citet{Engel1993} (EV) GGA and the \citebyname{Tran2009} modified 
Becke-Johnson (TB-mBJ) have also provided some improvements to the band gap of materials. For TB-mBJ, while the 
band gaps are considerably improved, the effective masses are severely underestimated \cite{PhysRevB.83.195134}. 
In the case of the EV potential, the equilibrium lattice constants are far too large as compared to experiment and, as such, 
leads to an unsatisfactory total energy \cite{PhysRevB.50.7279,Singh2010}.

The theoretical underestimations of band gaps and other energy eigenvalues have been ascribed to the inadequacies of 
density functional potentials for the description of 
ground state electronic properties of semiconductors \cite{Maus2002,Charlton2000,Padilla1997}. 
Also, other methods \cite{PhysRevLett.105.196403,RevModPhys.68.13, Bechstedt2009} that 
entirely go beyond the density functional theory (DFT) 
do not obtain the correct band gap values of most semiconductors 
without adjustment or fitting parameters \cite{Kim2010,Ahn2006}. 
This unsatisfactory situation is a key motivation for our work.

Stoichiometric STO has an experimental, indirect band gap of 3.20 - 3.25 eV at room temperature \cite{Bickel1989, Charlton2000, Ikeda1999}. 
Theoretical calculations using several techniques have led to band gaps of SrTiO$_{3}$ in the ranges 
1.71 to 2.2 eV for LDA and GGA \cite{Wang2001, piskunov2004, Jiangni2006}, 1.87 to 3.63 eV 
for Hybrid DFT \cite{Heifets2006, Eglitis2004, piskunov2004}, and 
value as high as 11.97 eV for the Hatree-Fock (HF) method \cite{piskunov2004}. 

In this paper, we present a simple, yet robust, and ab-initio method, based on self consistent 
solutions of the pertinent \textit{system} of equations 
\cite{Zhao1999, Bagayoko2007, Ekuma2011, Bagayoko2005}, that correctly predicts band gap values and 
related electronic properties of SrTiO$_{3}$ rigorously, from first 
principle, within the LCAO-GGA formalism. We also compute the structural, elastic and optical properties of SrTiO$_{3}$.   

The rest of this article is organized as follows. After this introduction in 
section ~\ref{sec:intro}, the computational methods and details are given in 
section ~\ref{sec:formalism}. The results of our self-consistent calculations are presented and discussed in 
section ~\ref{sec:results}. We then summarize and conclude in section ~\ref{sec:summary}. 

\begin{figure}
 \raisebox{-8cm}{{\includegraphics*[trim = 40mm 48mm 40mm 50mm, clip,totalheight=0.30\textheight, width=2.8in]{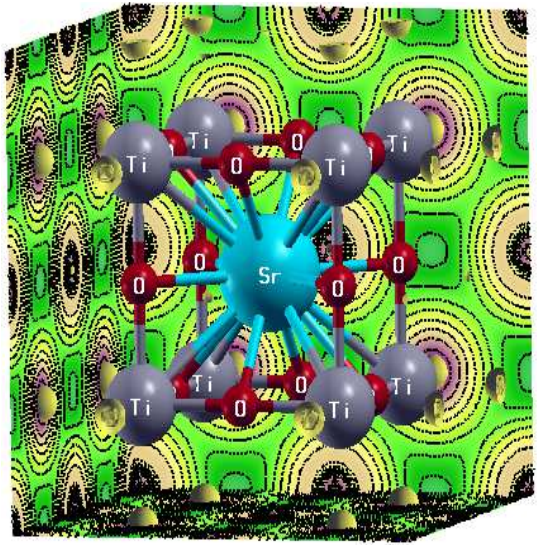}}}
\caption{(Color online). The iso-surface cubic unit cell of SrTiO$_{3}$ with the 
iso-lines, at lattice parameter of 3.91 \AA{}.} 
\label{fig:sro_isosurface}
\end{figure}\hspace{-1.0cm}      

\section{Methods and Computational Details}
\label{sec:formalism}
In the ground state, STO has the simple cubic (O$_{h}^{1}$ - Pm$\bar{3}$m) perovskite structure\cite{ICSD2011}, 
with Sr atom sitting at the origin, Ti at the body center and three oxygen atoms at the three 
face centers \cite{Wyckoff1963} (see Fig.~\ref{fig:sro_isosurface}). We used the room temperature 
experimental lattice constant of 3.91 \AA{} \cite{ICSD2011,Wyckoff1963}. 

\begin{figure}
\includegraphics*[totalheight=0.27\textheight, width=2.9in]{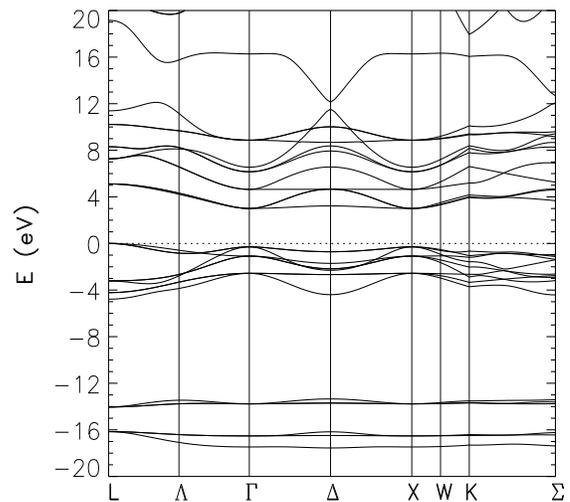}
\caption{Calculated, band structure of c-SrTiO$_{3}$ at the experimental lattice 
constant of 3.91 \AA{} as obtained with the optimal basis set using PW - GGA. The horizontal, dashed line indicates 
the position of the Fermi energy (E$_{F}$) (-0.10872 eV) which has been set equal to zero.} 
\label{fig:sro_bnd}
\end{figure} 

Our ab initio, self consistent, nonrelativistic calculations employed a linear combination 
of atomic orbitals (LCAO) formalism and a generalized gradient approximation (GGA) potential. 
One may argue that relativistic effects 
are important for the description of SrTiO$_{3}$. As was noted by 
\citebyname{Marques2003}, relativistic effects are only 
important for the description of the high-$\kappa$ dielectric band structure. 
Their calculated, relativistic and non-relativistic band structure 
for SrTiO$_{3}$, up to an energy of 5 eV for the valence and conduction bands, 
respectively, are almost identical. Consequently, we 
expect a negligible relativistic correction for the band gap of SrTiO$_{3}$.

The distinctive feature of our 
calculations, the use of the Bagayoko, Zhao, and Williams (BZW) method, has been extensively described 
in the literature \cite{Zhao1999, Bagayoko2007, Ekuma2010, Ekuma2011, Bagayoko2005, Ekumab2011, Bagayoko2004}. 
This method has been shown to lead to accurate ground state properties of 
many semiconductors: c-InN \cite{Bagayoko2004}, w-InN \cite{Bagayoko2005}, w-CdS 
\cite{Ekuma2011}, c-CdS \cite{ibid2011}, rutile-TiO$_{2}$ \cite{Ekumab2011}, 
AlAs \cite{Jin2006}, GaN, Si, C, RuO$_{2}$ \cite{Zhao1999}, 
and carbon nanotubes \cite{Zhao2004}. 

Instead of assuming that a single trial basis set will yield the correct ground state charge density of the solid, 
the BZW method entails a minimum of three self-consistent calculations with basis sets of different sizes and generally 
with different polarization functions, i.e., p, d, and f functions. The correct ground state is the one where all the 
occupied energies are at their minima. In practice, up to seven self-consistent 
calculations have been performed for some materials e.g, wurtzite ZnO \cite{Franklin}. The computations begin with a 
relatively small basis set that should not be smaller than the minimal basis set. The latter is defined as one that is 
just large enough to account for all the electrons in the atomic or ionic species present in the solid. The preliminary, 
self-consistent calculations of the electronic properties of the species provide the wave functions that serve as input in 
the solid state calculations.   

The first, self consistent calculation for the solid is performed with this 
small basis set (Calculation I) that is subsequently 
augmented with one orbital for the next self-consistent calculation (Calculation II). 
Depending on the angular symmetry of the added orbital, 
the size of the new basis set is larger than that of the previous one by 2, 6, 10, or 14 for $s$, $p$, $d$, and f functions, 
respectively. The occupied energies from calculations I and II are compared graphically and numerically. For the first 
two calculations, we found these occupied energies to be different for all the solids we have studied 
to date \cite{Ekumac2011,Bagayoko2008}, including SrTiO$_{3}$. The basis set for calculation II is then augmented in 
order to carry out self-consistent calculation III. Again, the occupied energies from calculations II and III are compared. 
This process of augmenting the basis set and of performing self-consistent calculations is continued until the occupied 
energies from a calculation, say $\textit{N}$, are found to be identical to their corresponding ones from calculation ($\textit{N+1}$), 
within computational uncertainties that are less than 50 meV. This perfect superposition of occupied energies from two 
consecutive calculations identifies the basis set for Calculation $\textit{N}$ as the optimal one, i.e, 
the smallest basis set that yields the lowest, occupied 
energies of the system. The attainement of this minima signifies that this basis set 
is verifiably complete for the description of the ground state of 
the system. Larger basis sets that include the optimal one do not 
lower any occupied energies from their values obtained with the optimal basis set. 

As explained elsewhere \cite{Ekuma2011,Zhao1999}, these larger basis sets do not 
lead to any changes in the ground state charge 
density or the Hamiltonian. However, larger basis sets that include the optimal one often lead to a lowering of some 
unoccupied energies. This lowering of some unoccupied energies cannot be ascribed to a 
physical interaction included in the Hamiltonian. Up to 
the optimal basis set, changes in the basis sets lead to changes in the charge density, the potential, and the Hamiltonian. 
Hence, changes in occupied and unoccupied energies, for self-consistent calculations with basis sets up to the optimal one, 
can be ascribed to a physical interaction embedded in the Kohn-Sham Hamiltonian. The system of equations in DFT 
totally determines changes in the occupied states. It also determines, at least in part, 
low-laying unoccupied states that are  interacting 
with the occupied ones, up to the optimal basis set. For example, in wurtzite InN, the calculated dielectric functions 
agree with their corresponding experimental ones up to 5.5-6.0 eV \cite{Jin2007}. 
Given that only direct transitions were taken into 
account in our dielectric functions calculations, this agreement indicates that the low-laying unoccupied 
bands were correctly determined. For larger basis sets that include the optimal one, the extra lowering of 
some \textit{unoccupied energies} is a direct consequence of the Rayleigh theorem 
\cite{Ekuma2010,Ekuma2011,Ekumab2011,Ekumac2011}. This theorem states that when an eigenvalue equation 
is solved with basis sets I and II, with set II larger than I and including I entirely, then the 
eigenvalues obtained with set II are lower than or equal to their corresponding ones obtained with basis set I 
\cite{Ekuma2011,Ekumab2011}.      

The above process entails iterations for the equation giving the ground state charge density, with the iterations for the 
Kohn-Sham equation carried out for each choice of the basis set. Given that iterations for the Kohn-Sham (KS) equation 
involves the charge density (CD) equation, one could conclude that a single trial basis set calculation solves 
both equations self consistently. A problem with this view stems from the fact that any two such calculations, 
with different trial basis sets, lead to different, converged (i.e., self consistent) eigenvalues of the KS 
equation. The fundamental theorem of algebra suffices to guarantee that the two sets will be different if 
the basis sets have different numbers of basis functions utilized in the expansions. 
The question then arises as to which of the two sets of eigenvalues of the KS equation 
provides the physical description of the system under study. To answer such a question definitively 
and from first principle, the BZW method follows the process described above. Upon reaching 
the optimal basis set, not only the charge density, the potential, and the Hamiltonian 
no longer change (i.e., they have converged), but also the resulting, self-consistent 
eigenvalues of the KS equation have reached their respective minima, for the occupied states. 
In our understanding, to solve the system of equation self-consistently means obtaining converged 
eigenvalues (attainable with most arbitrary basis set) but also $\textit{occupied eigenvalues}$ that have 
reached their respective minima (attainable with BZW method). 

In the above sense, the BZW method solves self-consistently not 
just the Kohn-Sham equation, but also the equation giving the ground state charge density in terms of the wave functions of 
the occupied states. Further, in his Nobel lecture \cite{Kohn1999}, Kohn noted 
the ``density optimal'' feature of the wave functions from correctly performed DFT calculations while those for the Hartree 
Fock approach are ``total-energy optimal.'' Without a constrained search for the converged ground 
state, it is quite difficult to infer the basis set that yields the correct ground state charge 
density \cite{Levy1982,Levy1979}. This point becomes clearer by noting that the reorganization of 
the cloud of valence electrons is drastically different for atomic or ionic species as compared to molecules or 
solids. For instance, \textit{single trial basis set calculations cannot make up for any deficiency in 
the angular symmetry of the 
functions, irrespective of the degree of convergence of the iterations of the Kohn-Sham equation}. By correct 
ground state charge density, we mean the charge density 
that leads to the minima of all the occupied energies. 

In this work, we utilize the electronic structure package from the Ames Laboratory 
of the US Department of Energy (DOE), Ames, Iowa \cite{Harmon1982}. We employ the generalized 
gradient approximation (GGA) potential given by \citebyname{Perdew1992}. We utilize sets of 
even tempered Gaussian functions with exponents from 0.12 to 10$^{5}$ to form the atomic wavefunctions. 
There are 15, 15, and 13 $\textit{s}$, $\textit{p}$, and $\textit{d}$ orbitals, respectively, for Sr, 
while 17, 17, and 15 $\textit{s}$, $\textit{p}$, and $\textit{d}$ orbitals, respectively, 
are used for both Ti and O. The charge fitting error using the Gaussian functions in the atomic calculation is 
about 10$^{-4}$. Since the deep core states are fully occupied and are inactive chemically, the charge densities of the deep 
core states were kept the same as in the free atom. However, the core states of low binding energy 
were still allowed to fully relax, along with the valence states, in the self consistent calculations.  
The orbitals employed in the self-consistent calculations are between paranthesis for 
Sr (3$\textit{d}$ 4$\textit{s}$ 4$\textit{p}$ 4$\textit{d}$ 5$\textit{s}$), Ti 
(3$\textit{p}$ 3$\textit{d}$ 4$\textit{s}$) and O (2$\textit{s}$ 2$\textit{p}$ 3$\textit{s}$), 
including some that are unoccupied in the free atoms or ions. 
These unoccupied orbitals are included in the self-consistent LCAO calculation to allow the 
restructuring of the electronic cloud, including possible polarization, in the crystal environment. 

The Brillouin zone (BZ) integration for the charge density in the self consistent procedure is 
based on 56 special k points in the irreducible Brillouin zone (IBZ). The computational error for the valence 
charge is 5.3 x 10$^{-5}$ eV per valence electron. 
The self consistent potential converged to a difference of 10$^{-5}$ after several tenths of iterations. 
The energy eigenvalues and eigenfunctions are 
then solved at 161 special k points in the IBZ for the band structure. A total of 152 weighted k points, 
chosen along the high symmetry lines in the IBZ of SrTiO$_{3}$, are used to solve for the 
energy eigenvalues from which the electron density of states (DOS) are calculated using the 
linear analytical tetrahedron method \cite{Lehmann1972}. The partial density of states (pDOS) and the effective 
charge at each atomic sites are evaluated using the Mulliken charge analysis procedure \cite{Mulliken1955}. 
We also calculated, the equilibrium lattice constant $a_{o}$, the bulk modulus ($B_{o}$), 
the associated total energy and the electron and hole effective masses in different directions.

In calculating the lattice constant, we utilize a least square fit of our data to the  
Murnaghan's equation of state \cite{Murnaghan1944,Murnaghan1995}. The lattice constant 
for the minimum total energy is the equilibrium one. 
The bulk modulus ($B_{o}$) is calculated at the equilibrium lattice constant.

The dielectric function $\varepsilon (\omega) = \varepsilon_{1} (\omega) + i \varepsilon_{2} (\omega)$  
can be calculated once the electronic wave functions and energies are known. The imaginary part of the 
dielectric function $\varepsilon_{2} (\omega)$, 
from the direct interband transitions, is calculated using the Kubo-Greenwood formula \cite{Bocquet1996}: 
\begin{widetext}
\begin{equation}
 \varepsilon_{2} (\omega)
=
\frac{8 \pi^{2} e^{2}}{3 m^{2} \hbar \omega^{2} \Omega} \sum_{k} \sum_{nl} | \langle \psi_{kn} (r) | P | \psi_{kl} (r) \rangle |^{2} f_{kl} [1 - f_{kn} ] \delta (\epsilon_{kn} - \epsilon_{kl} - \hbar \omega)
\end{equation}
\end{widetext}
where $\hbar \omega$ is the photon energy, $P$ = $ - i \hbar \triangledown$ is the momentum 
operator, $\Omega$ is the volume of the 
unit cell, $\psi_{kn} (r)$ and $\psi_{kl} (r)$ are the initial and final states, 
respectively, $f_{ki}$ is the Fermi distribution function for the $i_{th}$ 
states, and $\epsilon_{ki}$ is the energy of the electron in the $i_{th}$ state. 
The real part of the dielectric function, $\varepsilon_{1} (\omega)$, is obtained from 
the well-known Kramers-Kronig (KK) relation,
\begin{equation}
 \varepsilon_{1} (\omega)
=
1 + \frac{2}{\pi} M \int_{0}^{\infty} \frac{\varepsilon_{2} (\omega^{\prime}) \omega^{\prime}}{\omega^{\prime 2} - \omega^{2}} d \omega^{\prime},
\end{equation}
where $M$ indicates the principal value of the integral \cite{Ching1990}. 
The real part of the optical conductivity, $Re[\sigma (\omega)]$, follows from above as
\begin{equation}
Re[\sigma (\omega)]) = {\frac{\omega}{4 \pi} } \varepsilon_{2} (\omega)
\end{equation}

\begin{figure*}
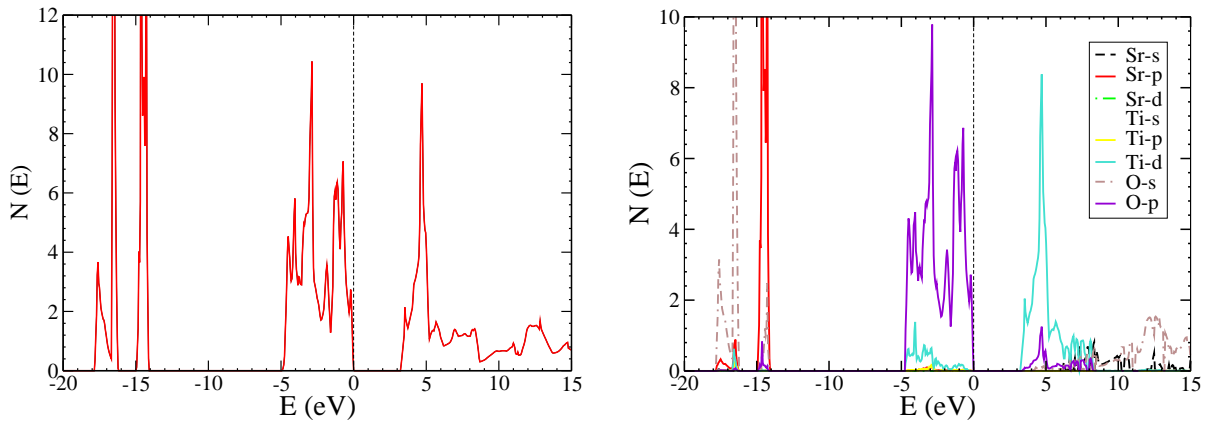

  \begin{center}
    \begin{minipage}[t]{0.470\linewidth}
      \raisebox{-5cm}{{\includegraphics[clip,angle=0,width=3.0in]{SRO_dos.eps}}}
    \end{minipage}\hspace{-0.3cm}
    \begin{minipage}[t]{0.470\linewidth}
      \raisebox{-5cm}{{\includegraphics[clip,angle=0,width=3.0in]{SRO_pdos.eps}}}
    \end{minipage}
\caption{\label{fig:sro_dos}(Color online). (a) Calculated, density of states (DOS) of c-SrTiO$_{3}$ at the experimental lattice constant of 
3.91 \AA{} as obtained with the optimal basis set using PW - GGA. (b) Calculated, partial density of states (pDOS) of c-SrTiO$_{3}$ at 
the experimental lattice constant of 3.91 \AA{} as obtained with the optimal basis set using PW - GGA. The vertical, dashed line 
indicates the position of the Fermi energy (E$_{F}$) which has been set equal to zero.}
\end{center}
\end{figure*}

\section{Results and Discussion}
\label{sec:results}
\begin{figure}
  \begin{center}
    \begin{minipage}[t]{0.50\linewidth}
      \raisebox{0cm}{{\includegraphics[trim = 20mm 43mm 35mm 40mm,totalheight=0.24\textheight, width=2.7in]{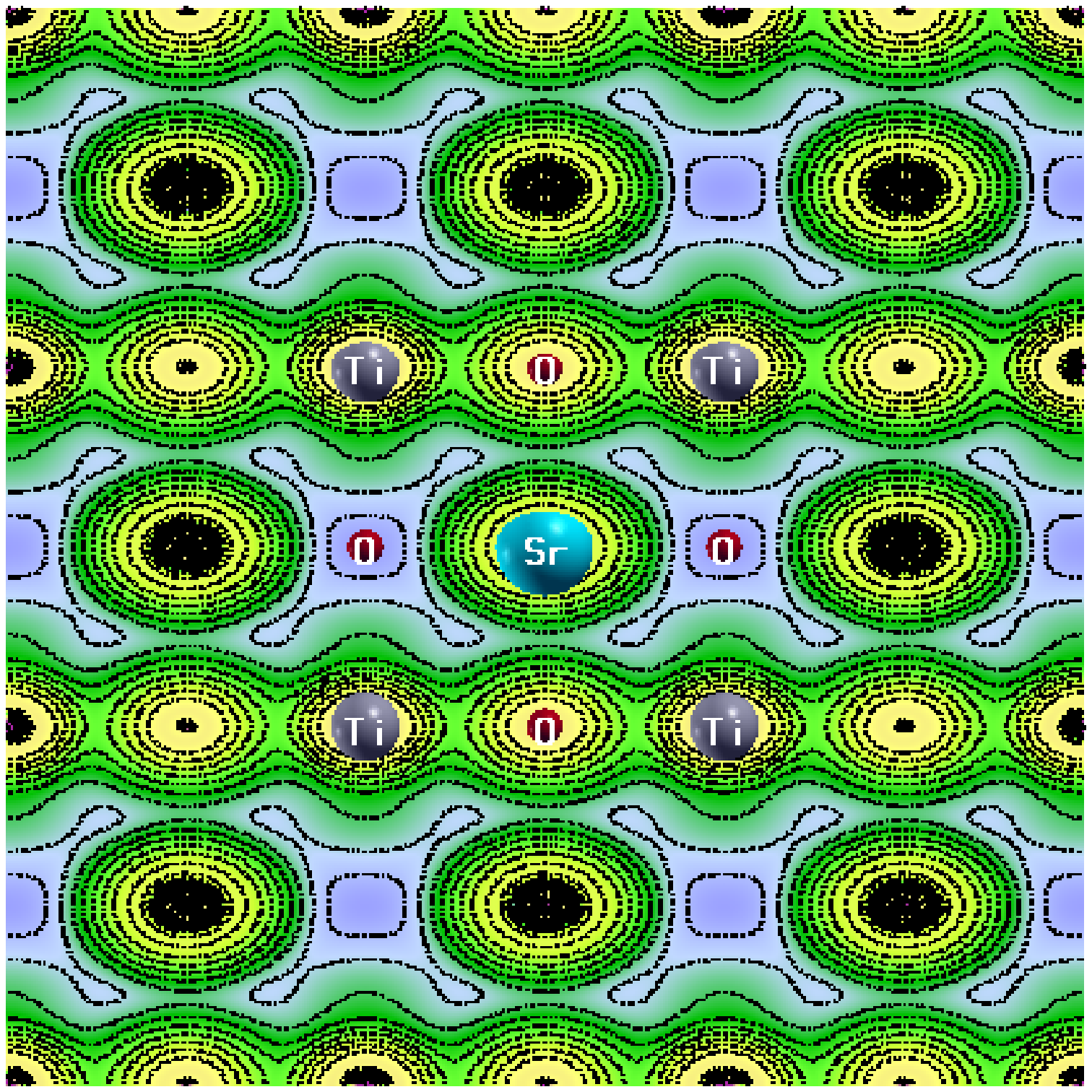}}}
    \end{minipage}\hspace{0cm} 
\quad\quad\quad\quad\quad
    \begin{minipage}[t]{0.30\linewidth}
      \raisebox{3.5cm}{{\includegraphics[clip,angle=0,totalheight=0.1\textheight,width=1.0in]{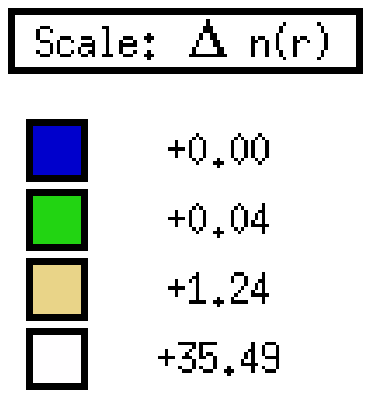}}}
    \end{minipage}
\caption{\label{fig:sro_dos100}(Color online). The contour plot of the 
electron charge density of c-SrTiO$_{3}$ as obtained with 
the optimal basis set using PW - GGA. The center atom is Sr with Ti atoms at the corners and O atoms in the middle of the 
size of the parallelogram around Sr. $\Delta$ $n(r)$ is the variation of the 
electron charge density as a function of distance away from an atomic site. A logarithmic scale 
has been used.}
\end{center}
\end{figure}

\begin{figure}
\includegraphics*[totalheight=0.24\textheight, width=2.9in]{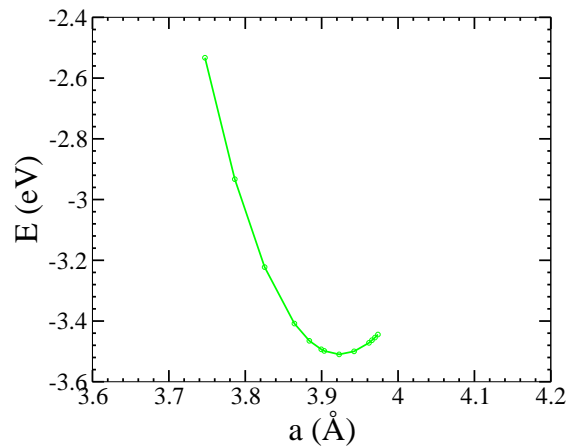}
\caption{(Color online) Calculated, total energy per unit cell as a function of 
the lattice constant of c-SrTiO$_{3}$, as obtained with the optimal basis set using PW - GGA. The calculated 
equilibrium lattice constant is 3.92 \AA{}.} 
\label{fig:sro_eos}
\end{figure}

The results of the electronic structure computations are given in 
Figs.~\ref{fig:sro_bnd} to~\ref{fig:sro_dos100}. Figure ~\ref{fig:sro_eos}
depicts the calculated total energy of STO, while Fig.~\ref{fig:sro_e} shows the optical spectra obtained using 
the optimal basis set from the electronic structure computations. 
Figures~\ref{fig:sro_isosurface} and~\ref{fig:sro_dos100} 
have been drawn using the xcrysden \cite{Kokalj2003}. 
We discuss the electronic structure, low laying 
conduction bands, and the effective mass in ~\ref{subsec:electronic}. 
The DOS is discussed elaborately in~\ref{subsec:dos}. 
The structural properties are presented in~\ref{subsec:structural}, while \ref{subsec:optic}  
deals with the calculated, optical properties.

\subsection{The Electronic Structure, Band Gap, Low-energy Conduction Band and Effective Mass}
\label{subsec:electronic}
The electronic structure of the valence and the low energy conduction states determine the band gap and other 
important properties of materials. Table~\ref{table:sro_comp} shows that our 
ab initio, first principle method yielded an indirect band gap of 3.24 eV at the experimental lattice 
constant of 3.91 \AA{} (see Fig.~\ref{fig:sro_bnd}) and an indirect band gap of 3.21 eV 
at the calculated equilibrium lattice constant of 3.92 \AA{}. This table contains several, previous results from calculations 
using LDA, GGA or hybrid potentials. Table~\ref{table:sro_eig} contains the calculated energies at some high symmetry points 
in the BZ. These energies are provided for future comparison with experimental and theoretical findings. Our calculated band 
structure (see Fig.~\ref{fig:sro_bnd}) resembles that of the parent TiO$_{2}$ system. 
Fig.~\ref{fig:sro_bnd} also shows that the top of the valence band is at the $L$ point. 

The effective mass is one of the main factors determining the transport properties, 
the Seebeck coefficient, and
electrical conductivity of materials. The calculated 
electron effective masses at the bottom of the conduction band along 
the $\Gamma$ - $L$, $\Gamma$ - $X$, and $\Gamma$ - $K$ 
directions, respectively, are 0.68 - 0.81, 0.44 - 0.59, and 0.51 - 0.66 while 
the calculated hole effective masses at the top of the valence band, along 
the $\Gamma$ - $L$, $\Gamma$ - $X$, and $\Gamma$ - $K$ 
directions, respectively, are 0.64 - 0.83, 1.22 - 1.27, and 0.96 - 1.02 (all in units of the electron mass). 
The observed anisotropy and the ranges of effective masses confirmed the earlier 
observations of Mattheiss and co-workers \cite{Mattheissa1972, Mattheissb1972}. Our calculated 
effective masses are in excellent agreement with the detailed effective mass values as reported 
by \citet{Mattheissa1972}[and references therein] and the relativistic computational results of \citebyname{Marques2003}. 

\begin{table}[ht]
\caption{Comparison of our calculated band gap values with other theoretical and experimental ones, 
for c-SrTiO$_{3}$. Our calculations show that c-SrTiO$_{3}$ has an indirect band gap from $L$ to $\Gamma$ points. 
Exactly the same band gap value is found from $L$ to $X$ points. All the band gaps are indirect unless 
otherwise indicated with ($\textit{D}$).}   
\centering                          
\begin{tabular}{c | c |c}            
\hline\hline                        
 & Computational Method & E$_{g}$ (eV) \\ [0.5ex]   
\hline                              
GGA & GGA - BZW (Present work) & 3.24 \\               
&(with equilibrium lattice constant) & 3.21 \\
&PP - PWGA & 1.97\footnotemark[1] \\
&PP - PBE & 1.99\footnotemark[1] \\
&FP - LAPW & 1.80\footnotemark[2] \\
&PP & 1.60\footnotemark[3] \\
&FLAPW & 1.78\footnotemark[4] \\
&TB - LMTO & 1.40 (D)\footnotemark[5] \\
&FB - LMTO & 2.20 (D)\footnotemark[6] \\
\hline
LDA    
&PP & 2.04\footnotemark[1] \\
&PP - PW & 1.79\footnotemark[13] \\
&LMTO - ASA & 1.80\footnotemark[7] \\
&PP & 1.71\footnotemark[8] \\
&PP - PW & 1.79\footnotemark[13] \\
&OLCAO & 1.45\footnotemark[14] \\
\hline
HF & PP & 11.97\footnotemark[1] \\
\hline
Hybrid DFT & B3PW - LCAO & 3.63\footnotemark[9] \\
&PP - BLYP & 1.94\footnotemark[1] \\
&PP - B3LYP & 3.57\footnotemark[1] \\
&PP - P3PW & 3.63\footnotemark[1] \\
&LCGO - B3PW & 3.70\footnotemark[10] \\
\hline                              
SA & FP - LAPW & 1.87 - 3.25\footnotemark[11] \\
\hline
Experiment & NA & 3.10 - 3.25\footnotemark[12] \\ [1ex]         
\hline\hline
\end{tabular}
\footnotetext[1]{Ref.~\cite{piskunov2004}.}
\footnotetext[2]{Ref.~\cite{Wang2001}.}
\footnotetext[3]{Ref.~\cite{Jiangni2006}.}
\footnotetext[4]{Ref.~\cite{Marques2003}.}
\footnotetext[5]{Ref.~\cite{Saha2000}.}
\footnotetext[6]{Ref.~\cite{Ahuja2001}.}
\footnotetext[7]{Ref.~\cite{Guo2003}.}
\footnotetext[8]{Ref.~\cite{Kim2010}.}
\footnotetext[9]{Ref.~\cite{Eglitis2004}.}
\footnotetext[10]{Ref.~\cite{Heifets2006}.}
\footnotetext[11]{Ref.~\cite{Cai2004}.}
\footnotetext[12]{Ref.~\cite{Wyckoff1963}.}
\footnotetext[13]{Ref.~\cite{Kimura1995}.}
\footnotetext[14]{Ref.~\cite{Mo1999}.}
\label{table:sro_comp}          
\end{table}

There is a significant O$_{2p}$ -- Ti$_{3d}$ hybridization in the valence bands. As shown in Fig.~\ref{fig:sro_bnd}, 
the valence bands of SrTiO$_{3}$ can be divided into three distinct groups: the upper, intermediate, and lower 
groups of valence bands occupy the energy ranges of 0 to -5.80 eV, -14.2 to -14.78, and -16.62 to -17.80 eV, respectively. 
The upper VB bands are made up of nine bands with a bandwidth of 5.80 eV, in 
agreement with X-ray photoemission spectroscopy (XPS) values of 5 - 6 eV \cite{Zollner2000} 
and the GGA results of \citebyname{Jiangni2006}. They are formed by the hybridization 
between O 2$\textit{p}$ and Ti 3$\textit{d}$, with very little contribution from the 
Ti 3$\textit{p}$ and Sr 4$\textit{p}$ (see Fig.~\ref{fig:sro_dos}(b)). 
Two of the bands at the $\Gamma$ point are triply degenerate: $^1 {\Gamma}_{15}$ (-2.75 eV) and $^{2} \Gamma_{15}$ (-0.25 eV), 
while the third band $\Gamma_{25}$ (-1.23 eV) is non-degenerate (see Table~\ref{table:sro_eig}). 
Immediately below the upper VB, 
a group of triply degenerate bands emanating from the hybridization between Sr 4$\textit{p}$, 
O 2$\textit{s}$ and O 2$\textit{p}$, with little dispersion at 
$^{3} \Gamma_{15}$. This group is located at -14.78 eV. 
The lowest lying VB bands are semi core like bands formed mainly due to 
the hybridization between O 2$\textit{s}$, and Sr 4$\textit{p}$ with very little bonding 
coming from Ti 4$\textit{s}$ and Ti 3$\textit{p}$. 
They are located at $^{1} \Gamma_{1}$ (-17.80 eV) and $^{2} \Gamma_{1}$ (-16.62 eV). Our calculated position of 
Sr 4$\textit{p}$ between - 14.78 to - 17.80 eV is in agreement with the XPS measurement of \citet{Battye1976} 
which placed it at 16.50 eV below the Fermi surface (E$_{F}$). Also, \citet{Board1970} and \citet{Battye1976} 
measured the average position of the O 2$\textit{s}$ to be about 17 eV below E$_{F}$. Our calculated value of 
-17.85 eV in Fig.~\ref{fig:sro_dos}(a)
is close to the experimental one. In particular, our result does not 
underestimate this core state position as is often the case in GGA calculations.

The conduction bands (CB), immediately above the Fermi level (low energy conduction band), are 
dominated by threefold degenerate Ti 3$\textit{d}$ $t_{2g}$ orbitals 
which hybridize with O 2$\textit{p}$ and O 2$\textit{s}$. 
The two-fold degenerate Ti 3$\textit{d}$ $e_{g}$ states have some hybridization 
from all other orbitals except Ti (4$\textit{s}$ and 3$\textit{p}$), Sr (4$\textit{p}$ and 3$\textit{d}$). 
The energy eigenvalue in the lowest conduction bands, at the $X$ point, are practically the same as that at 
the $\Gamma$ point, resulting in the observed, minimal dispersion in the conduction band between $\Gamma$ and $X$ points. 
This feature is apparent from our energies at the high symmetry points in Table II. 
At the $\Gamma$ point, energies associated with the lowest-laying conduction bands are: $^{1} \Gamma_{25^\prime}$ (3.24 eV), 
$^{1} \Gamma_{12}$ (4.84 eV), $^{2} \Gamma_{12}$ (6.10 eV), $^{3} \Gamma_{1}$ 
(6.75 eV) and $^{2} \Gamma_{25^{\prime}}$ (8.92 eV). The calculated, low energy 
conduction bands in Fig.~\ref{fig:sro_bnd} are quite different from those of previous studies. 
Figure~\ref{fig:sro_bnd} shows that the lowest conduction bands are degenerate 
at the $\Gamma$ and $X$ points along the [100] and equivalent directions. The electronic structure in 
Fig.~\ref{fig:sro_bnd} was calculated using the experimental lattice constant. 
We further examined whether or not the position of 
the shallow minimum in the lowest conduction band depends on the value of the lattice constant by using several 
values of the lattice constant around the experimental one. Even though, the band gap value changed from 
3.26 to 3.17 eV, there was no appreciable change in the depth of the shallow minimum in the lowest conduction band. 
We recall that the 
gap is 3.21 eV for our calculated, equilibrium lattice parameter.

\begin{table}[ht]
\caption{Eigenvalues (eV), along high symmetry points, for c-SrTiO$_{3}$, as obtained with the experimental lattice 
constant of 3.91 \AA{}. The Fermi energy of - 0.12188 eV is set to zero in the table. The energy eigenvalues at 
$\Gamma$ and $X$ points are found to be almost identical.}   
\centering                          
\begin{tabular}{c |c | c | c}            
\hline\hline                        
$L$ & $\Gamma$ & $X$ & $K$ \\ [0.5ex]   
\hline                              
-32.06 & -32.03 & -32.03 & -32.07 \\
-32.06 & -32.03 & -32.03 & -32.04 \\
-32.06 & -32.03 & -32.03 & -32.03 \\
-16.19 & -16.62 & -16.62 & -16.65 \\
-16.19 & -16.62 & -16.62 & -16.49 \\
-14.96 & -14.67 & -14.67 & -14.61 \\
-14.96 & -14.67 & -14.67 & -14.47 \\
-14.96 & -14.67 & -14.67 & -14.33 \\
-4.94 &  -2.75 & -2.75 & -3.92 \\
-4.32 & -2.75 & -2.75 & -3.43 \\
-4.32 & -2.75 & -2.75 & -3.09 \\
-3.58 & -1.23 & -1.23 & -1.68 \\
0 & -0.25 & -0.25 & -1.33 \\
0 & -0.25 & -0.25 & -1.04 \\
0 & -0.25 & -0.25 & -0.73 \\
5.22 & 3.24 & 3.24 & 4.06 \\
5.22 & 3.24 & 3.24 & 4.12 \\
5.22 & 3.24 & 3.24 & 4.33 \\
8.65 & 4.82 & 4.82 & 5.38 \\
10.92 & 10.28 & 10.28 & 13.42 \\
10.92 & 10.28 & 10.28 & 14.47 \\ 
\hline\hline
\end{tabular}
\label{table:sro_eig}          
\end{table} 

\begin{table*}[ht]
\caption{Experimental and theoretical lattice constants a (in \AA{}) for c-SrTiO$_{3}$ along with 
the calculated values of the bulk modulus (in GPa).}   
\centering                          
\begin{tabular}{c | c | c | c}            
\hline\hline                        
 & Computational Method & a (\AA{}) & B (GPa) \\ [0.5ex]   
\hline                              
GGA & BZW - LCAO (Present work) & 3.92 & 183.45 \\               
&PP - PWGA & 3.95 (3.93) & 167 (195)\footnotemark[1] \\
&PP - PBE & 3.94 (3.93) & 169 (195)\footnotemark[1] \\
&PW & 3.95 (3.88) & 167 (194)\footnotemark[2] \\
&PBE & 3.95 & 167 (194)\footnotemark[3] \\
&''   & 3.91 (3.82) & 210.21 (252.92)\footnotemark[4] \\
\hline
LDA & PP & 3.86 & 214 (215)\footnotemark[1] \\ 
&LAPW & 3.86 & 204 (176)\footnotemark[3] \\
&FLAPW &  3.95 & 167\footnotemark[3] \\
&''  & 3.93 (3.87) & 207.28 (227.63)\footnotemark[4] \\
&PP - PW & 3.87 & 194\footnotemark[5] \\
&OLCAO & 3.93 & 163\footnotemark[6] \\
\hline
HF & PP & 3.92 (3.93) & 219 (211)\footnotemark[1] \\
&PP & 3.98 (3.92) & 208.85 (206.68)\footnotemark[4] \\
\hline
Hybrid DFT & PP - BLYP & 3.98 & 164\footnotemark[1] \\
&PP - B3LYP & 3.94 & 177 (187)\footnotemark[1] \\
&PP - P3PW & 3.90 (3.91) & 177 (186)\footnotemark[1] \\
\hline                             
Experiment & NA & 3.89 - 3.92\footnotemark[7] & 174 - 183\footnotemark[8] \\ [1ex]         
\hline\hline
\end{tabular}
\footnotetext[1]{Ref.~\cite{piskunov2004}.}
\footnotetext[2]{Ref.~\cite{Bottin2003}.}
\footnotetext[3]{Ref.~\cite{Tinte1998}.}
\footnotetext[4]{Ref.~\cite{piskunov2000}.}
\footnotetext[5]{Ref.~\cite{Kimura1995}.}
\footnotetext[6]{Ref.~\cite{Mo1999}.}
\footnotetext[7]{Ref.~\cite{ICSD2011, Wyckoff1963, Hellwege1969, Mitsui1982}.}
\footnotetext[8]{Ref.~\cite{Wyckoff1963, Hellwege1969, Mitsui1982}.}
\label{table:sro_lat}          
\end{table*}

\subsection{Densities of States, Electron Distribution and Chemical Bonding}
\label{subsec:dos}
Figures~\ref{fig:sro_dos}(a) and~\ref{fig:sro_dos}(b) exhibit the total (DOS) and related partial (pDOS) densities of states, 
respectively. Figure~\ref{fig:sro_dos100} shows the contour plot of the 
distribution of the electron charge density of SrTiO$_{3}$. As can be seen from 
Fig.~\ref{fig:sro_dos100}, the electron density of SrTiO$_3$, away from the atomic sites, does not 
have a spherically symmetric distribution. Further, the bonding between Ti and O is covalent, 
due to Ti-3$d$ and O-2$p$ hybridization, unlike in the case 
of Sr and O that is ionic. The bond length of Sr -- O is 2.76 \AA{}, 
with a minimum charge density of $\sim$ 0.19 $e$/\AA{}$^3$, while Ti -- O bond length is 1.95 \AA{}, with 
a charge density of $\sim$ 0.63 $e$/\AA{}$^3$. The experimental bond lengths for Sr -- O and 
Ti -- O are 2.76 and 1.96 \AA{}, respectively, \cite{Kuroiwa2003,Jauch:sh5025} with 
corresponding charge densities of 0.2 
and 0.67 -- 0.90 $e$/\AA{}$^3$, in that order \cite{Ikeda1998,Friis:xc5016,Jauch:sh5025,Kuroiwa2003}. 
The charge density distribution around the O atom with respect to the horizontal 
Ti -- O -- Ti line is elongated in the direction 
along the Ti -- O covalent bond in agreement with room temperature experimental charge density distribution of 
SrTiO$_3$ reported by \citet{Ikeda1998}. This anisotropic charge density distribution 
is ascribed to the rotational mode of the 
Ti -- O$_6$ octahedron by experiment \cite{Jauch:sh5025,Kuroiwa2003,Abramov:sh0063}.

\begin{figure*}
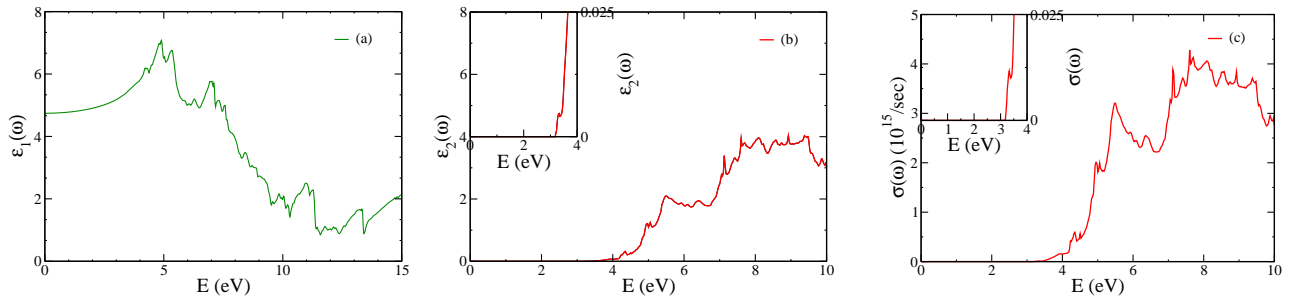

  \begin{center}
    \begin{minipage}[t]{0.30\linewidth}
      \raisebox{-5cm}{{\includegraphics[clip,angle=0,width=2.1in]{SRO_e1.eps}}}
    \end{minipage}\hspace{-0.3cm} 
\hfil
    \begin{minipage}[t]{0.30\linewidth}
      \raisebox{-5cm}{{\includegraphics[clip,angle=0,width=2.1in]{SRO_e2.eps}}}
    \end{minipage}
\hfil
    \begin{minipage}[t]{0.30\linewidth}
      \raisebox{-5cm}{{\includegraphics[clip,angle=0,width=2.1in]{SRO_cond.eps}}}
    \end{minipage}
\caption{\label{fig:sro_e}(Color online). (a) Calculated, dispersive part, $\varepsilon_{1} (\omega)$, of the dielectric function 
of c-SrTiO$_{3}$ as obtained with the optimal basis set using PW - GGA.  
(b) Calculated, absorptive part, $\varepsilon_{2} (\omega)$, of the dielectric function 
of c-SrTiO$_{3}$ as obtained with the optimal basis set using PW - GGA.  
(c) Calculated, optical conductivity, $\sigma (\omega)$, 
of c-SrTiO$_{3}$ as obtained with the optimal basis set using PW - GGA.  
As per the inserts (see Figs.~\ref{fig:sro_e}(b) and ~\ref{fig:sro_e}(c)), the band absorption edge is at 3.24 eV. 
In all cases, spectra have been calculated without any broadening.}
\end{center}
\end{figure*}

From our calculated DOS (see Fig.~\ref{fig:sro_dos}(a)), it can be inferred that the onset of absorption is quite sharp 
and it starts at about 3.24 eV. It exhibits a fine structure at 3.6 eV, with a shoulder at 4.50 eV. 
This picture is consistent with the experimental results of \citebyname{Cohen1968} and the theoretical findings of 
\citebyname{Perkins1983} of a relatively sharp absorption edge in the optical measurement of SrTiO$_{3}$. 
In the calculated DOS of the low laying conduction bands, sharp peaks appear at 4.70 eV, 5.72 eV, and 7.05 eV. 
Relatively broad peaks are found at 8.33 eV, 10.97 eV and 12.75 eV. Our computed peaks are in basic 
agreement with experimental findings of \citebyname{Cardona1965} and \citebyname{Braun1978}. For the valence bands DOS, 
we calculated peaks at -0.20 eV, -0.72 eV, -1.14 eV, -1.83 eV, -2.86 eV, -4.06 eV, -4.50 eV, 
-14.30 eV, -4.61 eV, and -17.53 eV. The peaks in the valence band DOS are all sharp. 
Our calculated electronic structure is in agreement with scanning transmission electron microscopy, 
vacuum ultraviolet spectroscopy, and spectroscopic ellipsometry measurement of \citebyname{Van2001}.

Our calculated band gap value of 3.24 eV, from $L$ to $\Gamma$, is practical the same as the experimental one. 
In general, other theoretical calculations obtained values 
that are up to 1.1 eV smaller. The source of the small band gap values was believed to be 
due to the pushing up of the top of the valence band dominated by Ti 3$\textit{d}$ 
and O 2$\textit{p}$ states to higher energies \cite{Pertosa1978}. 
According to our findings, it rather appears to be the extra lowering of the conduction bands that produces GGA (or LDA) 
band gaps that are more than 1.1 eV smaller than the experimental values, if LCAO type computations 
do not search for and utilize an optimal basis set. Such a basis set is verifiably converged for the description of 
occupied states \cite{Ekuma2011,ibid2011,Bagayoko2004}. We recall that Kohn and Sham \cite{Kohn1965}, in their original paper, 
explicitly stated the need to solve self consistently the pertinent \textit{system of equations} defining LDA. 
The BZW method, as explained 
above, rigorously solves the system of equations in the sense explained above. 
Single trial basis set calculations also involve both the KS and charge density equations. The major 
difference resides in the fact that these calculations do not generally entail changes in the basis functions 
beyond those of the expansion coefficients.  

\subsection{Structural Properties}
\label{subsec:structural}
The total energy versus the lattice constant data are shown in Fig.~\ref{fig:sro_eos}. 
The data fit well to the Murnaghan equation of 
state (EOS). The calculated equilibrium lattice constant is 3.92 \AA{} and the bulk modulus, $B_{o}$, is 183.45 GPa. 

The experimently reported lattice constants are in the range 3.89 to 3.92 \AA{}
\cite{ICSD2011, Wyckoff1963, Hellwege1969, Mo1999} and the bulk modulus lays in the range 174 to 183 GPa
\cite{Wyckoff1963, Hellwege1969, Mo1999}. 
In Table~\ref{table:sro_lat}, we show our calculated equilibrium lattice constant and bulk 
modulus in comparison with experimental and other theoretical results. Both our calculated 
lattice constant and bulk 
modulus agree well with corresponding, experimental ones, respectively.

\subsection{Optical Properties}
\label{subsec:optic}
The plot of the dispersive ($\varepsilon_{1} (\omega)$) and absorptive ($\varepsilon_{2} (\omega)$) dielectric functions are 
shown in Figs.~\ref{fig:sro_e}(a) and~\ref{fig:sro_e}(b), respectively, while the optical 
conductivity ($\sigma (\omega)$) profile is in Fig.~\ref{fig:sro_e}(c). 
All reported spectra have been calculated without any broadening and, may have more features than experimental ones. 
Our calculated, dielectric spectra are in good agreement with the experimental 
measurements\cite{Cardona1965, Mitsui1982}. The calculated optical spectra only included the 
direct inter band transitions. The fundamental absorption edge E$_{o}$, which is also a 
measure of the optical gap, was found to be 3.24 eV from the calculation, as per the insert of Fig.~\ref{fig:sro_e}(b). 
Our computed direct gap of 3.43 eV is in agreement with the experimental 
value of 3.40 eV\cite{Cardona1965, Mitsui1982}. Our calculated  
$\varepsilon_{1} (\omega = 0)$ at zero frequency equals 4.75 
(cf. Fig.~\ref{fig:sro_e}(a)). It compares well with the experimental 
value of 4.92 measured by \citebyname{Braun1978}. Our dielectric 
spectra resemble that of BaTiO$_{3}$ of \citebyname{Bagayoko1998}. 
This observation also holds for the data of \citebyname{Cardona1965} and \citebyname{Baurerle1978}. Both the 
experimental and our calculated results show that the direct optical gap is larger than the smallest indirect band gap. 

Figure~\ref{fig:sro_e}(c) shows the optical conductivity $\sigma (\omega)$ of SrTiO$_{3}$. 
As per the insert, it also shows that the fundamental absorption edge starts at 3.24 eV. 
The positions of the peaks (without any rigid shift) are in agreement 
with experimental findings.

\section{Summary and Conclusion}
\label{sec:summary}
We have performed first principle, ab initio calculations of the electronic, structural, elastic, and 
optical properties of bulk SrTiO$_{3}$ in the cubic phase using GGA potential and the BZW method. 
Our calculated results, 
without any adjustment or corrections, show good agreement with experimental data.

The agreement of our calculated band gaps (3.21 and 3.24 eV) and electron effective masses with 
corresponding, experimental values is significant. Some calculations with adjustable parameters 
can lead to the correct band gap; but they generally do not yield the correct curvature of the 
conduction band around its minimum-as given by the electron effective masses. Similarly, 
the agreement between the peaks in the calculated density of states with corresponding, 
experimental ones denote the correct description of the relative location of the bands. 
This result is confirmed by our reproduction of the measured features of the dielectric functions, 
the imaginary part of which was obtained using only direct transitions between occupied and unoccupied bands. 
Our calculated equilibrium lattice constant (3.92 \AA{}) and bulk modulus (183.45 GPa) also agree 
with corresponding, experimental findings.    

\begin{acknowledgments}
This work was funded in part by the the National Science Foundation 
(Award Nos. 0754821, EPS-1003897, and NSF (2010-15)-RII-SUBR), the Department of 
the Navy, Office of Naval Research 
(ONR, Award No. N00014-04-1-0587). CEE wishes to thank Govt. of Ebonyi State, Nigeria. 
\end{acknowledgments}


\end{document}